\documentclass[aps,pre,groupedaddress,showkeys]{revtex4-1}
\usepackage{amsmath,amsfonts,multirow,graphicx,subfigure}
\usepackage{wrapfig,mathtools}
\begin{document}

% Use the \preprint command to place your local institutional report
% number in the upper righthand corner of the title page in preprint mode.
% Multiple \preprint commands are allowed.
% Use the 'preprintnumbers' class option to override journal defaults
% to display numbers if necessary
%\preprint{}

%Title of paper
\title{Dependency structure and scaling properties of financial time series are related}

% repeat the \author .. \affiliation  etc. as needed
% \email, \thanks, \homepage, \altaffiliation all apply to the current
% author. Explanatory text should go in the []'s, actual e-mail
% address or url should go in the {}'s for \email and \homepage.
% Please use the appropriate macro foreach each type of information

% \affiliation command applies to all authors since the last
% \affiliation command. The \affiliation command should follow the
% other information
% \affiliation can be followed by \email, \homepage, \thanks as well.
\author{Raffaello Morales}
%\email[]{raffaello.morales@kcl.ac.uk}
%\homepage[]{Your web page}
%\thanks{}
%\altaffiliation{}
\affiliation{Department of Mathematics, King's College London, The Strand, London, WC2R 2LS, UK}
\author{T. Di Matteo}
\affiliation{Department of Mathematics, King's College London, The Strand, London, WC2R 2LS, UK}
\author{Tomaso Aste}
\email[]{t.aste@ucl.ac.uk}
\affiliation{Department of Computer Science, University College London, Gower Street, London, WC1E 6BT, UK.}
%Collaboration name if desired (requires use of superscriptaddress
%option in \documentclass). \noaffiliation is required (may also be
%used with the \author command).
%\collaboration can be followed by \email, \homepage, \thanks as well.
%\collaboration{}
%\noaffiliation

%\date{\today}
\begin{abstract}
%Complexity in financial markets reveals its signature through a mingle of randomness and regularity, with patterns often concealed under apparently random behaviour. 
We report evidence of a deep interplay between cross-correlations hierarchical properties and multifractality of New York Stock Exchange daily stock returns. The degree of multifractality displayed by different stocks is found to be positively correlated to their depth in the hierarchy of cross-correlations. We propose a dynamical model that reproduces this observation along with an array of other empirical properties. The structure of this model is such that the hierarchical structure of heterogeneous risks plays a crucial role in the time evolution of the correlation matrix, providing an interpretation to the mechanism behind the interplay between cross-correlation and multifractality in financial markets, where the degree of multifractality of stocks is associated to their hierarchical positioning in the cross-correlation structure. Empirical observations reported in this paper present a new perspective towards the merging of univariate multi scaling and multivariate cross-correlation properties of financial time series. 
\end{abstract}

% insert suggested PACS numbers in braces on next line
% insert suggested keywords - APS authors don't need to do this
\maketitle
%\maketitle must follow title, authors, abstract, \pacs, and \keywords
\section{Introduction}
\begin{figure}[h!]
\includegraphics[width=0.8\textwidth]{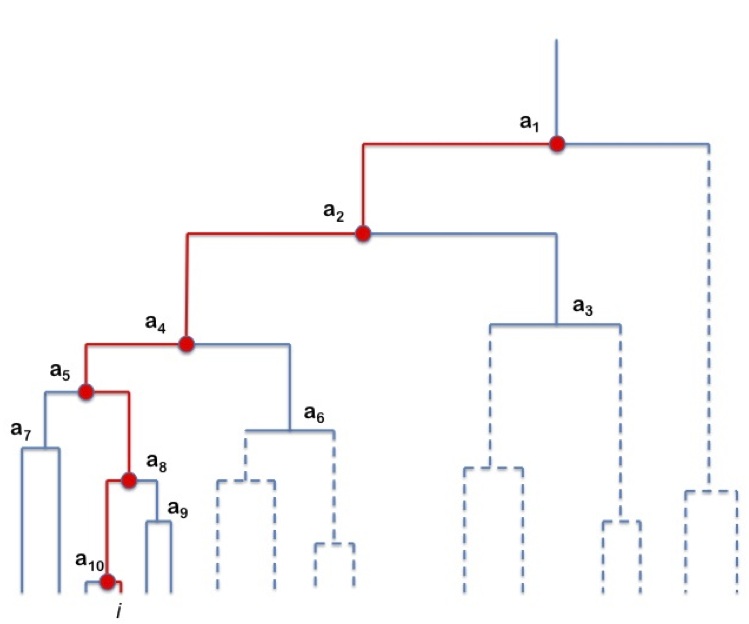}
\caption{{\bf Example of hierarchical structure.} The path highlighted in red is $\Gamma_{i,t}$, while the thick red bullets are the nodes $a_{m},\,\,m\in\gamma_{i,t}$, which correspond to the risks stock $i$ is exposed to. Note that the risk $a_{1}$ is common to all stocks, while other risks affect groups of stocks only. The dashed branches of the dendrogram indicate an arbitrary hierarchy. }
\label{fighierarchy}
\end{figure}

Financial markets dynamics is driven by forces that show their signatures ubiquitously through the complex behaviour of the price historical time series \cite{dacorogna2001introduction}. A major challenge is to seek connections between different aspects of complexity and to come up with some common mechanism able to explain these aspects coherently. There are two main elements that define the complexity of financial time series: the first is multifractality \cite{mandelbrot1997fractals,matteo2005long,di2007multi}, which is associated to the behavior of each single variable and the way it scales in time; the second is the structure of dependency between time series \cite{mcneil2005quantitative}, associated with the collective behavior of the whole set of variables. So far, these two manifestation of complexity have been investigated separately. In this paper we point out that -in fact- they are related and we propose a model that can reproduce the observed relationship.
\newline In spite of their intrinsic complexity, prices show remarkable regularities, which are commonly referred to as stylised facts \cite{mantegna2000introduction,chakraborti2011econophysics,cont2001empirical}. These empirical features are somehow the footprint of some underlying layout which distinguishes prices from purely random processes. Indeed, researchers have been progressively stranding away from the hypothesis of Brownian motion, originally proposed by Bachelier in 1900 \cite{bachelier1900theorie}, as a realistic model for the price evolution and more complex dynamics along with more refined models and mechanisms have been proposed over the past twenty years \cite{malevergne2006extreme}. For instance, prices increments at low frequencies (intra-day to weekly) are known to show non-Gaussian behaviour and other distributions, for example Student-t with degrees of freedom in the range $[3,5]$ have been shown to better fit financial returns \cite{bouchaud2003theory}. Another ubiquitous property not exhibited by random noise is the presence of persistence in the price increments, responsible for a phenomenon which is commonly referred to as volatility clustering \cite{mandelbrot1963variation}. Volatility tends to cluster particularly over turbulent periods, where large fluctuations are observed more frequently than in tranquil market periods. Both fat tails and time persistence are captured by a more general behaviour widely observed in financial markets across different asset classes: multifractality. Multifractality is a measure of the complexity of the process, but at the same time it reflects its regularities, mirroring some of the most relevant stylised facts. Many studies have investigated the multifractal behaviour of financial time series \cite{di2003scaling,barunik2010hurst,liu2007true,liu2008multifractality,lux2004detecting,lux2008markov,bouchaud2000apparent} and several models have been proposed in order to explain its phenomenology \cite{ding1993long,calvet2002multifractality,bacry2001multifractal,eisler2004multifractal,filimonov2011self}.  
\newline Further to the complex and regular properties of the single prices, financial markets also exhibit distinctive cross-dependency patterns, which are well captured in the framework of network theory \cite{caldarelli2007scale}. The complex dependency structure of a multivariate dataset can be mapped into a structure of nested hierarchies of correlations through clustering methods \cite{mantegna1999hierarchical,tumminello2005tool,tumminello2010correlation} which extract a hierarchy of nested clusters of stocks. The hierarchical organisation provides a very useful display of the hidden dependency patterns governing a set of stocks: it reveals which prices are more strongly correlated with respect to the rest of the market and it has been shown to match the intuition of explaining market moves in terms of few hierarchical factors \cite{tumminello2007hierarchically}. 
%In particular, the redundant information present in the correlation matrix of a multivariate data set can be filtered out through efficient methods which exploit only topological constraints on the network of correlations \cite{mantegna1999hierarchical,tumminello2005tool}. These methods turned out to be extremely useful both as filtering tools to retain relevant correlations and as means to detect hierarchical structures in the market \cite{tumminello2010correlation}. On this line, a very recent non deterministic approach has been devised in \cite{song2012hierarchical} which extracts a hierarchy of nested clusters of stocks starting from some topological properties of the filtered graph of correlations. The hierarchical organisation provides a very useful display of the hidden dependency patterns governing a set of stocks: it reveals which prices are more strongly correlated with respect to the rest of the market and it has been shown to match the intuition of explaining market moves in terms of few hierarchical factors \cite{tumminello2007hierarchically}. 
%\newline The interplay between complexity and regularity in financial markets therefore emerges from different aspects which can be related to distinct properties of the prices.
\newline In this article we investigate the interplay between two measures of complexity in the stock prices time series, namely their multifractality and their hierarchical order measured through the DBHT clustering algorithm \cite{song2012hierarchical}. The first measure reflects properties of the single time series and, as repeatedly shown in many works \cite{zhou2009components,kantelhardt2002multifractal}, is influenced by two main factors: (a) a broad unconditional distribution of the process increments and (b) the long-range non-linear time dependence of the increments. The second deals with the complex cross correlation structure of a multivariate set of time series and therefore necessarily reflects some forms of interaction between stock prices. Here we show that these two measures of complexity, associated with the irregularities present in the prices, are remarkably intertwined and show cogent correlations, especially for certain classes of stocks belonging to specific sectors of the market and to the set of clusters which best express the corresponding sectors. Starting from these observations, we postulate the existence of a hierarchical structure of risks which can be deemed responsible for both stock multivariate dependency structure and univariate multifractal behaviour. We then construct a model that reproduces these empirical observations, built on the hypothesis that a complex hierarchy of risk factors affects the stocks heterogeneously. We show that a simple assumption of nested multiplicative risks can explain the observed correlations between multifractality and hierarchical order, hence providing new insights into the underlying mechanism governing financial markets.
%\newline  The complex dependency structure of a multivariate dataset can be mapped into a structure of nested hierarchies of correlations through the DBHT clustering method \cite{song2012hierarchical}. Taking as input the correlation matrix of the dataset, after filtering it via the PMFG \cite{tumminello2005tool}, that is retaining only the correlations that respect planarity of the graph obtained from the correlation matrix, the method arranges the stocks into a hierarchical structure extracted from the topology of the graph.  
\newline The DBHT identifies clusters of stocks and generates a hierarchical organisations both intra-clusters and inter-clusters, which can be visualised by means of a dendrogram. While the intra-cluster hierarchy reveals the dependence between the different clusters, the inter-cluster hierarchy provides additional information about the nested organisation of stocks inside each cluster. We will refer to the junctions in the dendrograms as dendrogram nodes (or simply nodes) and to the hierarchical structure above (under) the cluster level as super (sub)-cluster hierarchy respectively. 
\newline The hierarchical structure produced by the DBHT provides a natural way to associate with each stock a hierarchical order, that is the number of  nodes above each stock along the path from the stock at the bottom of the tree to the top of the dendrogram. We will denote with $n$ the generic order and with $n_{i}$ the hierarchical order of a specific stock $i$, where $i=1,\dots,N$. In other words, $n_{i}$ measures how deep down the hierarchical tree lies a certain stock $i$. A schematic hierarchical structure is given in Figure \ref{fighierarchy}. The hierarchical tree $\Gamma_{i,t}$ is defined to be the set of nodes along the simple path $\Gamma_{i,t}$ from the stock to the top of the dendrogram, i.e. $\Gamma_{i,t}=\{a_{m}: m\in\gamma_{i,t}\}$, where we denote by $\gamma_{i,t}$ the set of positive numbers identifying the nodes above stock $i$. Note that there's only one such simple path for each stock $i$. Then the hierarchical order is defined as the cardinality of $\Gamma_{i,t}$, $n_{i}=\text{card}(\Gamma_{i,t})$. In the example reproduced in Figure \ref{fighierarchy}, for the stock labeled by $i$, we have $\gamma_{i,t}=\{1,2,4,5,8,10\}$, $\Gamma_{i,t}=\{a_{1},a_{2},a_{4},a_{5},a_{8},a_{10}\}$, i.e. the set of nodes denoted by the red dots and $n_{i}=6$. 
\newline In this paper we conjecture that the hierarchical order $n_{i}$ can be actually viewed as a measure of the riskiness of stock $i$, because of its positive dependence with multifractality. This entails that the cross-correlation properties of the multivariate dataset are somehow entangled with the stylised facts displayed by the univariate time series. In particular the hierarchical structure of correlation provides likewise a snapshot of the hierarchy of risks in the market. To the best of our knowledge, up to now no attempt to uncover this kind of underlying structure has been made in the literature.
%Note that, although stocks with higher order $n$ are, by construction of the hierarchy, those displaying the larger number of significant correlations with all other stocks, it is not obvious that this fact be reflected in the multi factor nature of the risk affecting the stocks. Indeed, it is not straightforward to show that correlation risk influence the complexity of the single stocks and, to the best of our knowledge, no attempt to uncover this kind of underlying structure has been made in the literature. If, on the contrary, the DBHT clustering provides, aside from the hierarchical structure of correlation, a snapshot of the hierarchy of risks in the market, then we would expect the heterogeneity of risks to be somehow correlated with other measures of complexity and hence riskiness. 
\section{Results}
\subsection{Cluster detection and sectors}
We have considered daily stock prices comprising the 342 most capitalised stocks continuously traded in the NYSE in the period 2-01-1997 to 31-12-2012. Data have been provided by Bloomberg. The dataset includes stocks from 9 different market sectors, according to the Bloomberg classification. The taxonomy of the stocks in the respective sectors is given in the Supplementary Material (SM), where we also report details on the clusters detected through DBHT clustering algorithm. We have chosen to perform the clustering through DBHT after having verified that it works better than other methods including the Single Linkage Cluster Analysis (SLCA) \cite{gower1969minimum} in recovering a well-diversified hierarchical structure: SLCA in fact tends to produce very large hubs of stocks that bias the correlation hierarchical order. 
\subsection{Correlation between multifractality and hierarchical order}
We have looked at the correlation between the hierarchical order and the multifractal properties of the stocks. 
%The finding of a positive (or negative) dependence between the two quantities strongly supports the idea that a multi-factor underlying structure of different risks be the driving-force of both correlations and multi-scaling behaviour of the stock prices \textit{per se}.
\newline As an indicator of degree of multifractality we have considered, as in previous studies \cite{morales2012dynamical, morales2013non}, the quantity
\begin{equation}
\Delta H(1,2)=H(1)-H(2),
\end{equation}
where $H(q)$ for $q=1,2$ is the generalised Hurst exponent computed from the linear scaling of the empirical q-moments (see Methods Section for more details). We have removed from the analysis all stocks whose multifractality cannot be statistically distinguished from zero, which correspond to weak multifractal behaviour and hence would not be relevant in this context. The benchmark value for multifractal stocks has been set  to $\Delta H(1,2)>0.015$ (see Methods). 
\begin{figure}[h!]
\includegraphics[width=0.8\textwidth]{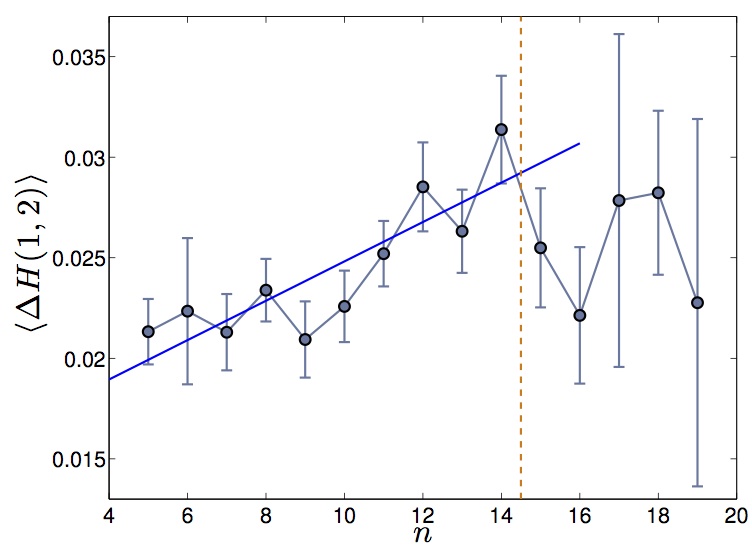}
\caption{{\bf Demonstration that multifractality and hierarchical order are positively correlated.} (Color online) We plot in blue circles with error bars the multifractal indicator $\langle \Delta H(1,2)\rangle$ averaged over the stocks sharing the same hierarchical order, against the hierarchical order $n$. The blue solid line is the linear fit over the averages, while the orange horizontal dashed line marks the limit up to which the increasing trend is observed. The error bars are the standard errors computed as $s/\sqrt{N}$, where $s$ is the standard deviation over the stocks having same hierarchical order. }
\label{figAllStocks}
\end{figure}
\newline We plot in Figure \ref{figAllStocks} the mean value $\langle \Delta H(1,2)\rangle$ with standard error $s/\sqrt{N}$ (with $s$ the standard deviation on the mean) for each observed hierarchical order on all stocks analysed (see Methods for details). We observe a positive dependence between the two variables up to $n=14$, followed by some noisier flat trend. The positive correlation between $\langle \Delta H(1,2)\rangle$ and $n$ is also confirmed by performing a t-test on the correlation coefficient validated with p-value $p=0.03$. All stocks with hierarchical order in the range $[5,14]$ (which accounts for $90\%$ of all stocks) exhibit multifractal properties increasing along with their depth in the hierarchy of correlations. On the other hand, the apparent saturation observed for orders larger than $14$ suggests that the hierarchical structure of cross-correlations may be responsible for the multifractal properties of the stocks only up to a certain order. Let us note in particular that the number of stocks found with hierarchical order $n>14$ is too small to allow any robust statistical conclusion, which is also the reason why standard errors in Figure \ref{figAllStocks} are very large for $n>14$. 
\newline We have investigated this relationship for other markets as well, namely: London Stock Exchange (LSE), Tokyo Stock Exchange (TSE) and Hong Kong Stock Exchange (HKSE). Results and details are reported in Sections 2-3 of the SM. The interdependence between multifractality and hierarchical cross-correlation order has been confirmed on data from LSE (see Figure 3 in SM). The two Asian markets show a much wider range of hierarchical order due to the appearance of one very large hub in the correlation network, which includes most of the stocks. This hub biases the evaluation of the proper hierarchical order. However, by detrending the time series we can remove the large hubs and, at least for TSE data and only for the small hierarchical orders, retrieve a similar behaviour to that observed in Figure \ref{figAllStocks} (see Figure 5 in Supplementary Material). As explained in detail in the SM though, detrending the series seriously affects the scaling properties of the data and the overall trend in the plots $\langle \Delta H(1,2)\rangle$ $vs$ $n$ is fundamentally flat. Nonetheless we mention (and report in Figure 4 in SM) that without detrending the series the positive dependence between $\Delta H(1,2)$ and $n$ is observed (for small $n$) in both TSE and HKSE data. 
%This results in a reduction of the database of around $20\%$ of its size (72 stocks out of 342). 
\newline We have also detected the same positive dependence between hierarchical order and multifractality on specific market sectors and on clusters found through DBHT clustering algorithm. We show in the top of Figure \ref{figFinInd} plots of the multifractality indicator versus the hierarchical order computed on stocks belonging to Financial and Industrial sectors. The black dots are values for single stocks whereas the red squares are the average multifractal indicators for each order. We also plot the best fit on the dots (thick blue line). Both sets of stocks show a very well defined positive correlation between the hierarchical order $n$ and the multifractality indicator $\Delta H(1,2)$, which is evident from the positive trend recovered in both examples. Again, the dependence is found to be significant with p-value $p=0.02$. 
\newline In Figure \ref{figFinInd} we also report the trends observed on two of the largest clusters found through the DBHT. The same positive correlation between multifractality and hierarchical order is observed on the clusters best identifiable with the corresponding sectors: the Financial sector has large components in clusters $2,4$ and $6$, whereas a large component of stocks from the Industrial sector is found in cluster $3$ (Table 2 in the SM).
\begin{figure}
\includegraphics[width=0.8\textwidth]{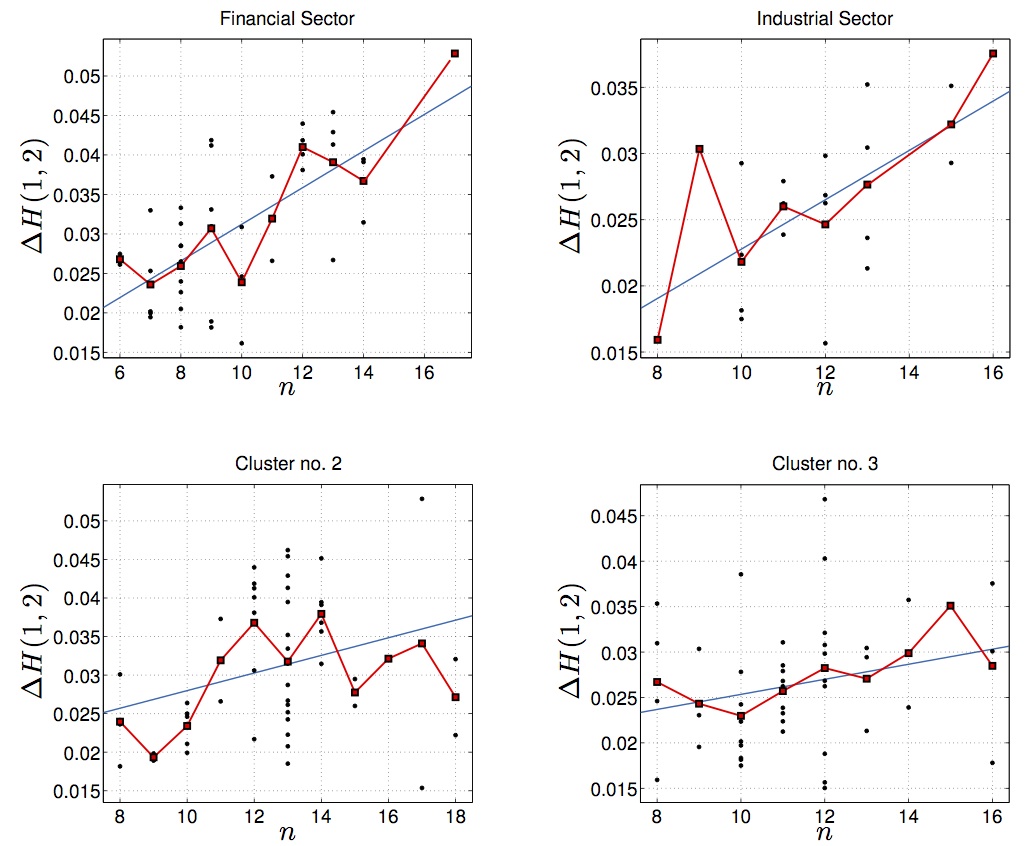}
\caption{{\bf Correlation between multifractality and hierarchical order in single sectors and clusters.} We plot in black dots $\Delta H(1,2)$ against the hierarchical order $n$ for stocks in the Financial (top left) and Industrial (top right) sectors and cluster 2 (bottom left) and cluster 3 (bottom right). In all plots the blue line is the best fit of the dots, while the red squares are the averages of $\Delta H(1,2)$ for each fixed order. }
\label{figFinInd}
\end{figure}
All stocks belonging to the largest clusters and the corresponding sectors that better match the clusters show a tendency to pair high multifractality with depth in the hierarchy of correlations (for other examples see SM). Moreover, the positive trends have been also observed when restricting the attention on sub clusters, i.e. groups of stocks below the cluster level in the hierarchy. More details about the analysis on sub clusters is reported in the SM. We also report in the SM a bootstrap study which retains only the stocks whose hierarchical order is validated after resampling with replacement the empirical time series. We find that the most robust hierarchical orders show even neater correlation with the corresponding multifractality indicators and this further analysis helps reinforcing the significance of the trends shown in Figure \ref{figFinInd}. 
\newline Although in all cases shown multifractality is found to depend on the hierarchical order, this dependence is likely to be non-linear. Let us consider for example the stocks in cluster $2$: the trend is clearly increasing for $n\in[9,14]$, then there is some oscillation around somewhat less than the value of multifractality attained for $n=14$ similarly to what observed on the complete set of stocks in Figure \ref{figAllStocks}.
\newline The correlation between multifractality and hierarchical order has been also detected dynamically in time. We have performed DBHT clustering in time on 50 overlapping time windows of length $752$ days. The hierarchical structure evolves significantly in time revealing time varying topological properties of the dendrograms. In particular we tracked the number of clusters $\mathcal{N}_{c,t}$ observed on the time window $t$. This gives a measure of how compact the hierarchy is: a large number of clusters corresponds to a hierarchy sprawled horizontally, while a small number of clusters corresponds to a hierarchical tree narrow and deep and therefore compact. The observed behaviour of $\mathcal{N}_{c,t}$ is reported in the left plot in Figure \ref{figgammat}. We observed a systematic decrease of $\mathcal{N}_{c,t}$ with time, with the trend being steeper in the period preceding the 2007-2008 financial crisis. Hence, the overall contraction of the market already reported in other studies \cite{onnela2003dynamic,aste2010correlation,onnela2003dynamics} corresponds to a contraction of the hierarchy, which reflects the fact that the market tends to show less heterogeneity in correlations. Note that to the shrinkage of the hierarchical structure corresponds an increase of the average correlation $\langle \rho\rangle_{t}$ in the market, as shown in the right plot in Figure \ref{figgammat}.
\begin{figure}
\includegraphics[width=0.8\textwidth]{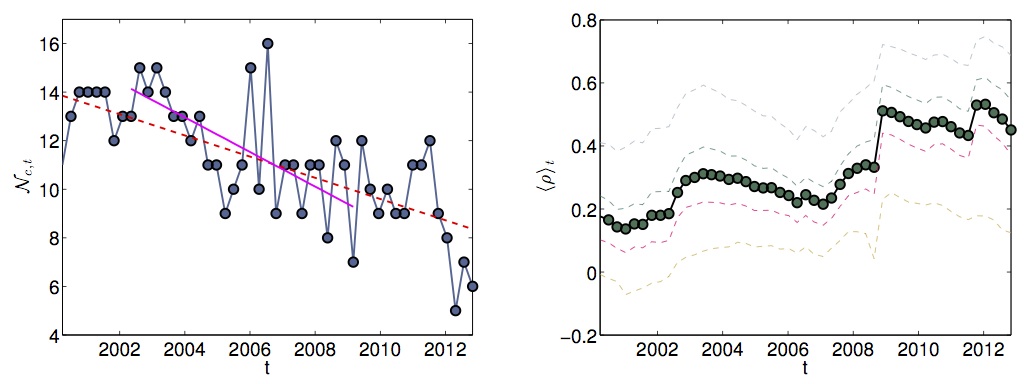}
\caption{\label{figgammat} {\bf Coalescence of the hierarchy in time and dynamical correlation.} (Left) The number of cluster $\mathcal{N}_{c,t}$ as a function of time is plotted in time in blue circles. The dashed red line is the best fit over the entire time period 2-01-1997 to 31-12-2012, while the magenta line is the fit over the shorter period preceding the 2007/2008 financial crisis September 2002 through November 2007. Standard errors on the circles are not reported as too small to be visible. (Right) Dynamical evolution of the average correlation (thick dots) in the same time period. The dashed coloured lines represent the $2.5\%,25\%,75\%,97.5\%$-quantiles, taken from the distribution of all the observed correlation coefficients.}
\end{figure}
Remarkably, the increasing coalescence of the hierarchical structure is followed through by an increase in both average multifractality $\langle \Delta H(1,2)\rangle_{t}$ and average hierarchical order $\langle n\rangle_{t}$, whose behaviours in time are reported in Figure \ref{figAVDH_n}. 
\begin{figure}
\includegraphics[width=0.8\textwidth]{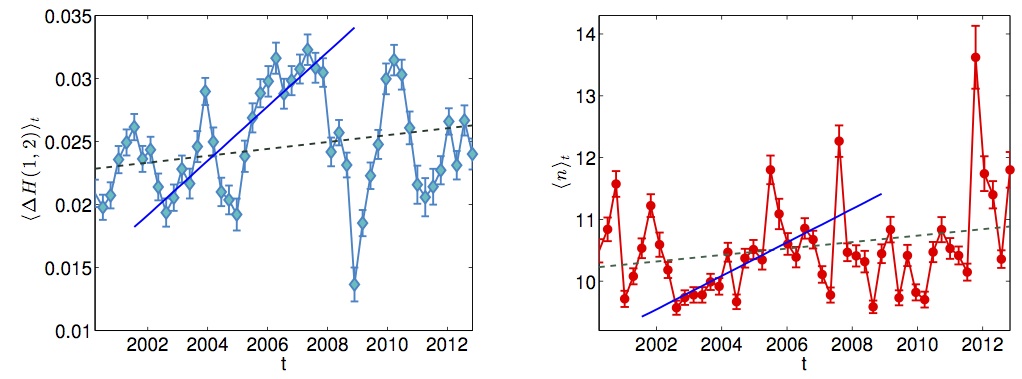}
\caption{\label{figAVDH_n}{\bf Average multifractality and hierarchical order in time.}  (Left) The blue diamonds are values of the average multifractality over 50 overlapping time windows. The dashed green line is the best fit over the entire time period 2-01-1997 to 31-12-2012, while the blue line is the fit over the shorter period preceding the 2007/2008 financial crisis, September 2002 through November 2007. (Right) The red circles are values of the average hierarchical order over 50 overlapping time windows. The dashed green line is the best fit over the entire time period 2-01-1997 to 31-12-2012, while the blue line is the fit over the shorter period preceding the 2007/2008 financial crisis, September 2002 through November 2007. In both plots the error bars are the standard mean error on the mean $s/\sqrt{N}$, with $s$ the standard deviation.}
\end{figure}
The increase in multifractality can be explained through the exposure of the prices to a larger set of risks, which seem to be well captured by the increase in the average hierarchical order. 
\newline The observed trends strongly suggest a deep interplay between the cross dependence patterns of the market and the multifractal properties of the stock prices. A natural interpretation is then to consider the hierarchical order as the number of different risks affecting each stock together with all other stocks sharing the same hierarchical structure. These observations give in fact a new insight into the mechanism responsible for the heterogeneity of multifractality in stock prices. Backed up by numerous empirical studies which have confirmed the prominent role of the fat tailed nature of the distribution of returns as the main source of observed multifractality \cite{zhou2009components,kantelhardt2002multifractal,barunik2012understanding,morales2013phd}, the observations presented here point in the direction of a mechanism of price formation where the \textit{measured} multifractal behaviour emerges from the collective action of many different risks, which are in turn captured in the market hierarchical structure. The combined action of many risks has the effect of swelling the tails of the returns distribution, triggering an increase in the measured degree of multifractality \cite{bouchaud2000apparent}. 
\subsection{A Multivariate Dynamical Hierarchical Model}
\begin{figure}
\includegraphics[width=0.8\textwidth]{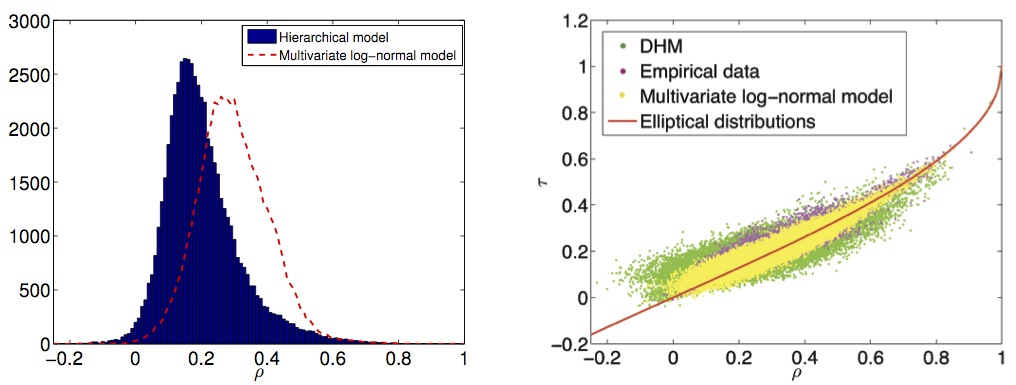}
\caption{\label{figdistRho} {\bf Modification in the correlation structure by means of the hierarchical model.} (Left) We plot in blue bars the histogram of the observed pair correlations on 342 DHM simulated time series, with hierarchical structure extracted from empirical data. The dashed red line is the histogram observed on a multivariate log-normal model, corresponding to the absence of any hierarchical structure. (Right) We plot Kendall $\tau$ against linear correlation $\rho$ for multivariate log-normally distributed synthetic time series (yellow dots), synthetic DHM time series (green dots) and empirical data (magenta dots), compared with the theoretical relation expected for elliptically distributed random variables (red thick line). }
\end{figure}

\begin{figure}
\includegraphics[width=0.8\textwidth]{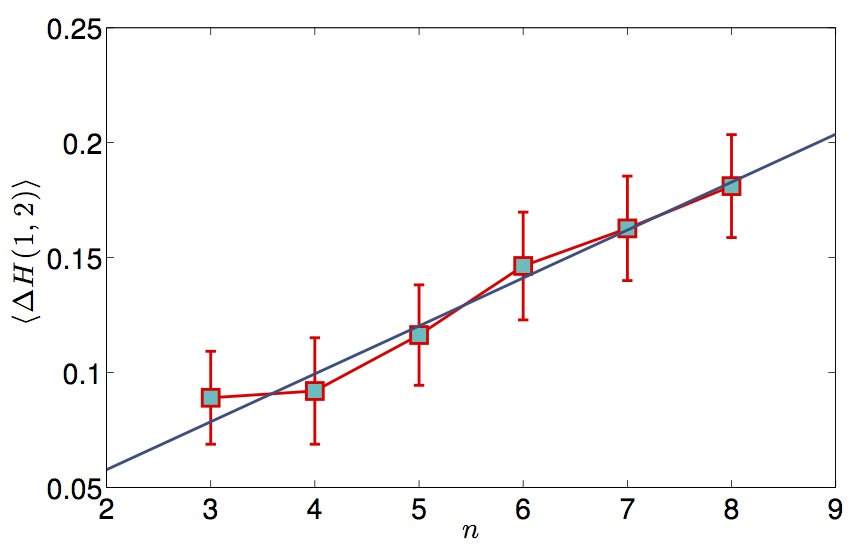}
\caption{\label{figDHnmodel} {\bf Demonstration that DHM time series with higher hierarchical order show larger multifractality.} We plot in squares the average values of $\Delta H(1,2)$ for each hierarchical order $n_{i}$. The averages are computed over $1000$ realisations of DHM simulations with the hierarchical structure extracted from empirical data and probabilities initialised randomly with values ranging in $[0,1]$. The error bars are standard deviations over the observed $\Delta H(1,2)$ for each hierarchical order. The blue line is the best fit on all the simulated stocks. }
\end{figure}

\begin{figure}
\includegraphics[width=0.8\textwidth]{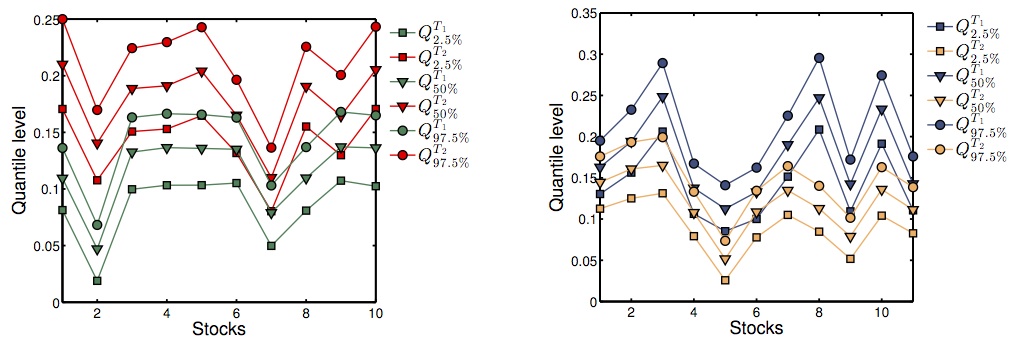}
\caption{\label{figDHquantiles} {\bf Time varying multifractality with two-regime hierarchical structure.} (Left) The plot shows the q-quantiles $Q_{q}^{T_{1}}$ and $Q_{q}^{T_{2}}$ for $q=\{2.5\%,50\%,97.5\%\}$ for the 10 DHM simulated time series whose hierarchical order $n_{T_{1}}<n_{T_{2}}$. Red lines correspond to p-quantiles in $T_{1}$ while black lines to p-quantiles in $T_{2}$. (Right) The q-quantiles $Q_{q}^{T_{1}}$ and $Q_{q}^{T_{2}}$ for $q=\{2.5\%,50\%,97.5\%\}$ for the 11 DHM simulated time series whose hierarchical order $n_{T_{1}}>n_{T_{2}}$. Blue lines correspond to q-quantiles in $T_{1}$ while orange lines to q-quantiles in $T_{2}$. The probabilities of the risks are initialised to values ranging in $[0.4,0.6]$. }
\end{figure}

In order to explain the mechanism underlying the observed link between multifractality and correlation hierarchy we introduce a dynamical hierarchical model (DHM), whose main novelty with respect to standard multivariate models \cite{mcneil2005quantitative} lies in the introduction of a perturbation term on the correlation matrix associated to its hierarchical structure. We model returns of stock $i$ as 
\begin{equation}
r_{i,t}=\epsilon_{i,t}\sigma_{i,t},
\end{equation}
where $\epsilon_{i,t}=(\boldsymbol{\epsilon}_{t})_{i}$ is a stationary multivariate Gaussian random variable, i.e. $\boldsymbol{\epsilon}_{t}\sim\mathcal{N}(0,\Sigma)$ with $\Sigma$ the covariance matrix and $\sigma_{i,t}$ is a volatility factor. Differently from usual multivariate models, $\sigma_{i,t}$ is not common to all stocks but depends explicitly on the hierarchical structure of the market. We suppose there is a volatility factor $x_{t}$ common to all stocks and then a latent hierarchical structure of risks $a_{m}$ with $m=1,\dots,N-1$, associated with the nodes of the dendrogram. The process $x_{t}$ is chosen as a stationary log-normal stochastic process autocorrelated in time, with the autocovariance function decaying as a power law, i.e. $\text{Cov}(x_{t}x_{t+h})\sim h^{-1}$, in line with a sweeping amount of studies on multi fractals in finance \cite{bacry2001multifractal,calvet2002multifractality}. For the arbitrary stock $i$ with hierarchical path $\Gamma_{i,t}=\{a_{m},\,\,m\in\gamma_{i,t}\}$ we define the volatility factor as
\begin{equation}
\sigma_{i,t}=x_{t}\prod_{m\in\gamma_{i,t}}e^{K_{m,t}}, 
\end{equation} 
where $K_{m,t}$ are Bernoulli random variables with probabilities $p_{m}$ associated to the nodes $a_{m}$ along the hierarchical tree $\Gamma_{i,t}$. Returns therefore take the form $r_{i,t}=\epsilon_{i,t} Y_{n,t}^{(i)} x_{t}$, where $Y_{n,t}^{(i)}=\prod_{m\in\gamma_{i,t}}e^{K_{m,t}}$. It is worth remarking that the hierarchical term $Y_{n,t}^{(i)}$ introduces a richer structure of dependence, where the topology of the risk organisation plays a crucial role in creating heterogeneous dependence that cannot be accounted for by standard multivariate models.  
\newline The correlation matrix $\rho$ of this model has entries $\rho_{ij}=\text{Corr}(r_{i}r_{j})=\text{Corr}(\epsilon_{i}\epsilon_{j}) \mathcal{F}_{ij}({\bf p};\Gamma_{i,t},\Gamma_{j,t})$, where (see SM for proof)
 \begin{equation}\label{PertFactor}
 \mathcal{F}_{ij}({\bf p};\Gamma_{i,t},\Gamma_{j,t})=\frac{\prod\limits_{l:a_{l}\in\Gamma_{i,t}\setminus\Gamma_{j,t}}\zeta_{1}(p_{l}) \prod\limits_{h:a_{h}\in\Gamma_{j,t}\setminus\Gamma_{i,t}}\zeta_{1}(p_{h})}{\left(\prod\limits_{l:a_{l}\in\Gamma_{i,t}\setminus\Gamma_{j,t}}\zeta_{2}(p_{l}) \prod\limits_{h:a_{h}\in\Gamma_{j,t}\setminus\Gamma_{i,t}}\zeta_{2}(p_{h}) \right)^{1/2}}.
 \end{equation}
In the last equation ${\bf p}$ denotes the set of probabilities involved in the relevant dendrogram, $\zeta_{1}(p)=p(e-1)+1$ and $\zeta_{2}(p)=p(e^2-1)+1$. $\rho_{ij}$ is therefore factorized into two contributions: the correlation coefficient of the multivariate random vector ${\boldsymbol \epsilon}_{t}$ and the hierarchical factor $\mathcal{F}_{ij}({\bf p};\Gamma_{i,t},\Gamma_{j,t})$. Although the expression of the perturbation factor $\mathcal{F}_{ij}({\bf p};\Gamma_{i,t},\Gamma_{j,t})$ may look rather daunting, it has a very simple interpretation. In both limit cases where the probabilities associated with the risks are all zero or all one, one has $\mathcal{F}_{ij}({\bf p};\Gamma_{i,t},\Gamma_{j,t})= 1$, which corresponds to the hierarchy effect being turned off. For all other possible values of the probabilities one has $\mathcal{F}_{ij}({\bf p};\Gamma_{i,t},\Gamma_{j,t})<1$. The correlation coefficient $\rho_{ij}$ is therefore shrunk by a highly diversified hierarchical structure but increases when the hierarchy coalesces, which is exactly what has been observed empirically and reported in Figure \ref{figgammat}. Note that both cases $p_{m}=0$, $\forall m=1,\dots,N-1$ and $p_{m}=1$, $\forall m=1,\dots,N-1$ correspond to the absence of a diversified hierarchical structure: in the first case indeed all nodes are switched off, leaving only the common factor to contribute to the volatility, in the second case all risks are active with probability one and thus there is no diversification in the hierarchical configurations of each stock. We show in the left plot in Figure \ref{figdistRho} the distribution of measured correlations on a multivariate synthetic DHM with probabilities $p_{m},\,\,m=1,\dots N-1$ uniformly distributed in the range $[0.1,0.4]$ compared to a multivariate log-normal model, which is recovered from the DHM when all $p_{m}$'s are one. One can see how the distribution in the case of $\mathcal{F}_{ij}({\bf p};\Gamma_{i,t},\Gamma_{j,t})\neq1$ is more skewed to the left. Increasing the values of the probabilities progressively shifts the median to the right (see SM). The limit $\mathcal{F}_{ij}({\bf p};\Gamma_{i,t},\Gamma_{j,t})=1$ corresponds to the case where the market is dominated by one single risk and the multi-branched hierarchical structure disappears. This scenario corresponds to the single factor log-normal volatility multivariate model, to which the DHM reduces when the correlation increases. The conjecture of this underlying hierarchical structure becomes even more appealing on account of a recent study \cite{chicheportiche2012joint} about the inadequacy of elliptical distributions to describe the stock returns dependency structure. The authors have shown that the single factor volatility seems to work quite well when the correlation coefficient between two stocks is large, but performs poorly for small correlations. Their observations fit well in the framework of our model, where the hierarchical factor is needed to explain small correlation. A comparison between multivariate log-normally distributed synthetic time series (yellow dots), synthetic DHM time series (green dots) and empirical data (magenta dots) is reported in the right plot in Figure \ref{figdistRho}, where we also plot the theoretical relation expected to hold between linear correlation and Kendall's $\tau$ for the class of elliptical distributions, $\tau=2/\pi\arcsin\rho$ \cite{lindskog2003kendall}. We can see that while the multivariate log-normal model cannot fully explain the real stocks departing from the theoretical relation, the DHM allows for a much broader dispersion and thus can better explain the empirical observations. 
The width of the dispersion about the theoretical relation depends on the probabilities $p_{m}$ and the green cloud of dots collapses onto the yellow one in the limit $p_{m}\to1$ or $p_{m}\to0,\,\,m=1,\dots,N-1$. 
\newline Overall one can say that the DHM allows for time varying correlation and different correlation regimes are ascribed to a time varying structural organisation of risks in the market. The non-stationary nature of correlations \cite{toth2006increasing,munnix2012identifying} is a very relevant issue in quantitative finance, and any sound portfolio strategy should take this feature into account \cite{livan2012non}. 
\newline Further to the time varying correlation patterns, the DHM is also capable of reproducing the observed correlation between hierarchical order and multifractality. The intuitive idea behind the mechanism is that high multifractality of the signals is dominated by large fluctuations rather than by a complex time dependence structure. Hence, although the time correlation structure of the common volatility factor stays the same, the upsurge in the number of observed extreme events can trigger multifractality to increase dramatically. As an example, we illustrate the properties of the model on a restricted multivariate data set of 25 stocks, whose hierarchical structure retrieved via DBHT are reported in the SM. The hierarchical order $n$ is found to vary in the range $[3,8]$. The behaviour of the average $\langle\Delta H(1,2)\rangle$ as function of the hierarchical order is shown in Figure \ref{figDHnmodel} for $1000$ simulations of DHM with the same hierarchical structure. We observe clearly that deeper stocks tend to have larger multifractality, as a result of the coordinate action of the many risks. Conversely, stocks with small hierarchical order show smaller multifractality in comparison. 
\newline The model also reproduces time varying multifractality along with a varying hierarchical structure. We consider a two-regime hierarchical structure extracted from empirical data by means of DBHT over two different time windows, labelled $T_{1}$ and $T_{2}$ respectively. We fix $T_{1}=T_{2}=2013$ days, corresponding to half the length of the empirical time series of returns. The two corresponding hierarchies are labelled $\mathcal{H}_{T_{1}}$ and $\mathcal{H} _{T_{2}}$. We plot in Figure \ref{figDHquantiles} the q-quantiles $Q_{q}^{T_{1}}$ and $Q_{q}^{T_{2}}$ for $q=\{2.5\%,50\%,97.5\%\}$ of the distribution of $\Delta H(1,2)$ measured on the simulated DHM time series with hierarchies $\mathcal{H}_{T_{1}}$ in the first tranche and $\mathcal{H} _{T_{2}}$ in the second. Comparing the hierarchical orders of the stocks in both tranches, we plot in different colours the quantiles on the two different time windows, both for stocks whose hierarchical order increases (left) and those whose hierarchical order decreases (right).
We observe a systematic shift of the distribution of  $\Delta H(1,2)$ towards larger values for those stocks whose hierarchical order increases and a shift towards smaller values for those whose hierarchical order decreases. This confirms further that the underlying structure of risks can be a good candidate to explain empirical observations presented above as well as empirical observations of time-varying multifractality \cite{morales2013non}. 
\newline Let us also mention that this model is capable of reproducing all the relevant statistical properties of asset returns. The autocorrelation function of the square returns simulated with probabilities in the range $[0,1]$ is found to decay as a power law with the lag $h$, specifically $C(r_{t}^2r_{t+h}^2)\sim h^{-\beta}$ with $\beta\in [0.3,0.6]$, in line with what observed on financial time series \cite{cont2001empirical}. The model also reproduces tails thicker than those accounted for by a simple log-normal volatility model \cite{bacry2008continuous}: the common volatility factor is in fact not sufficient to explain the values of the tail exponents $\alpha$ of financial returns probability distributions, which, as discussed for example in  \cite{bouchaud2003theory}, are found to be in the range $[1,5]$. The hierarchical structure instead can produce such small values of $\alpha$ (particularly $\alpha<2$) thanks to the action of the many hierarchical risks. The risks associated with the internal nodes increase sensibly the excess kurtosis, which, particularly for probabilities in the range $[0.5,1]$, assumes values comparable to those observed on empirical data.   
\section{Discussion}
We have shown that multifractality in financial markets is not independent from the hierarchical order measured through a correlation based clustering algorithm. We have detected that, within a reasonably large range of hierarchical orders, stocks deeper in the hierarchy of correlations are also those showing higher degree of multifractality. This result holds globally on the average multifractality measured on the entire market as well as locally for all major market sectors and the corresponding clusters detected by the DBHT clustering algorithm. This main empirical result has been also verified on a dataset from a different western market while it is less evident in markets showing very large hubs of correlated stocks (see example of the Japanese and Hong Kong market in SM). We have also shown that to the contraction of the market typically unfolding before a financial crisis corresponds a shrinkage of the hierarchical structure and an average increase of multifractality. Starting from these observations we have proposed a model which incorporates explicitly a time varying hierarchical structure of risks into the volatility factor. The presence of the hierarchical factor in the volatility results in a perturbation to the correlation structure of the multivariate time series which is shown to include also the empirical observations not accounted for by a standard multivariate model with one common volatility factor only. The latter is recovered in the limit of no hierarchy of risks, which corresponds to a very compact market where one single risk is enough to explain all correlations. The model has been shown to faithfully reproduce the correlation between multifractality and hierarchical order thus providing a new mechanism to explain the source of heterogeneity and time dependence of multifractality in financial markets, already reported in previous works \cite{morales2012dynamical,morales2013non}. 
\newline The relevance of the observations presented in this paper lays in the identification of a possible underlying mechanism where cross-correlation and univariate time series properties are merged together in a unified framework. 
\newline An interesting outlook is to devise a method to identify the underlying structure of risks conjectured in this paper in order to come up with a reliable calibration scheme which could be further used to develop portfolio strategies or asset allocation based on a perturbed correlation matrix.  
\section{Methods}
\subsection{Multifractality measure}
The multifractality proxy $\Delta H(1,2)$ is estimated via the generalised Hurst exponent (GHE) method \cite{di2007multi}. The GHE $H(q)$ is estimated from the scaling of the empirical q-moments $M(q,\ell)=\frac{1}{N-\ell+1}\sum_{t=0}^{N-\ell}|r_{t,\ell}|^q\sim \ell^{H(q)q}$, where we denote $r_{t,\ell}=\log p_{t+\ell}-\log p_{t}$ the log-return at time $t$ and scale $\ell$, with $p_{t}$ the asset price at time $t$. The exponent $H(q)$ is then estimated as average of several linear fits of  the relation $\log M(q,\ell)\sim qH(q)\log\ell$ with scale $\tau\in[1,\ell_{max}]$ and $\ell_{max}$ varied in the range $[5,19]$. We have considered as significantly multifractal only those stocks showing $\Delta H(1,2)>0.015$. This value has been chosen as the $95\%$ probability value of $\Delta H(1,2)$ on uniscaling processes. To obtain it we have simulated 1000 realisations of fractional Brownian motion with Hurst parameter $H$ randomly chosen in the range $[0.1,0.9]$ and then computed the value of $\Delta H(1,2)$ corresponding to (two side) $95\%$ probability. In our data base of 342 stocks we found 72 stocks excluded as weakly multifractal. 
\newline The average values of $\Delta H(1,2)$ for each different order $\bar{n}$ reported in Figures \ref{figAllStocks} and \ref{figFinInd} have been computed as $\langle \Delta H(1,2)\rangle_{\bar{n}}=\frac{1}{N_{\bar{n}}}\sum_{i:n_{i}=\bar{n}}\Delta H_{i}(1,2)$, where $N_{\bar{n}}$ is the number of stocks with order $\bar{n}$. Likewise the standard deviation is computed as $s=\sqrt{\frac{1}{N_{\bar{n}}-1}\sum_{i:n_{i}=\bar{n}}(\Delta H_{i}(1,2)-\langle \Delta H(1,2)\rangle_{\bar{n}})^2}$.  
\subsection{Dependency measure and DBHT}
The clustering is performed on the weighted graph associated with the correlation matrix of the empirical time series, whose entries are computed via the weighted Pearson estimator \cite{pozzi2012exponential}. We have implemented, in order to reduce the impact of remote events on present correlations, exponential weights $w_{t}=w_{0}\text{exp}\left(\frac{t-\Delta t}{\theta}\right)$ such that $w_{t}>0$ and $\sum_{t=1}^{\Delta t}w_{t}=1$. The parameter $\theta$ has been set to $\theta=\Delta t/3$ according to criteria previously established \cite{pozzi2012exponential}. This corresponds to $\theta=1342$ for the entire time window and $\theta=250$ for the moving windows of 752 days. 
\subsection{Model simulation}
The DHM time series have been simulated multiplying a multivariate Gaussian random vector $\epsilon_{t}$ with location parameter $\mu=0$ and covariance matrix $\Sigma$ times the volatility factor $\sigma_{t}$. The covariance matrix $\Sigma$ has been chosen to be the covariance matrix of the real data. The volatility factor $\sigma_{t}$ has been simulated multiplying a time correlated log-normal process $x_{t}$ times a hierarchical factor $Y_{n,t}^{(i)}$, whose components match the hierarchy detected through DBHT. The process $x_{t}$ has been simulated as $x_{t}=e^{\xi_{t}}$, with $\xi_{t}\sim\mathcal{N}(0,1)$. The autocorrelation function of $\xi_{t}$ has been chosen, according to the cascade picture in \cite{bacry2001multifractal}, as $\text{Cov}(\xi_{t}\xi_{t+h})=\lambda^{2}\log\frac{T}{1+h}$, where $\lambda$ and $T$ are two parameters, which, in all simulations reported in this paper, have been set to $\lambda=0.2$ and $T=800$. Other values have also been extensively tested, showing negligible effects on multifractality properties compared to the hierarchical factor. We have initialised the probabilities of the dendrogram nodes to random values uniformly drawn in different ranges, always checking that the choice would not bias the outcome of the simulation. In the multi-regime simulations, we have first computed the covariance matrices of the data $\Sigma_{T_{1}},\dots,\Sigma_{T_{d}}$ on different time windows $T_{1},\dots,T_{d}$ and then performed DBHT on each time window. The different hierarchical structures $\mathcal{H}_{T_{1}},\dots,\mathcal{H}_{T_{d}}$ recovered have been plugged into the model, keeping both $\epsilon_{t}$ and $x_{t}$ stationary on all the different tranches.   
\begin{acknowledgments}
The authors wish to thank Jean-Philippe Bouchaud and N. Musmeci for helpful discussions. TDM wishes to thank the COST Action TD1210 for partially supporting this work.
\end{acknowledgments}
\section{Author contributions}
All authors contributed equally to the manuscript writing the paper, revising it and preparing figures and tables.  
% Create the reference section using BibTeX:
%merlin.mbs apsrev4-1.bst 2010-07-25 4.21a (PWD, AO, DPC) hacked
%Control: key (0)
%Control: author (72) initials jnrlst
%Control: editor formatted (1) identically to author
%Control: production of article title (-1) disabled
%Control: page (0) single
%Control: year (1) truncated
%Control: production of eprint (0) enabled
%

\end{document}